\documentclass[twocolumn,showpacs,preprintnumbers,amsmath,amssymb]{revtex4}
\usepackage{graphicx}% Include figure files
\usepackage{dcolumn}% Align table columns on decimal point
\usepackage{bm}% bold math

\begin{document}

%\preprint{APS/123-MW}

\title{Dynamical instability and domain formation in a spin-1
Bose condensate}

\author{Wenxian Zhang, D. L. Zhou, M. -S. Chang, M. S.
Chapman, and L. You}

\affiliation{School of Physics, Georgia Institute of Technology,
Atlanta, Georgia 30332-0430}

%\affiliation{$^2$Institute of Theoretical Physics, The Chinese Academy of
%Sciences, Beijing 100080, China}

%\author{Wenxian Zhang, D. L. Zhou, M. -S. Chang, M. S. Chapman, and L. You}

%\affiliation{School of Physics, Georgia Institute of Technology, 837 State
%Street, Atlanta, Georgia 30332-0430}

\date{\today}

\begin{abstract}
We interpret the recently observed spatial domain formation in spin-1 atomic
condensates as a result of dynamical instability. Within the mean field theory,
a homogeneous condensate is dynamically unstable (stable) for ferromagnetic
(antiferromagnetic) atomic interactions. We find this dynamical instability
naturally leads to spontaneous domain formation as observed in several recent
experiments for condensates with rather small numbers of atoms. For trapped
condensates, our numerical simulations compare quantitatively to the
experimental results, thus largely confirming the physical insight from our
analysis of the homogeneous case.
\end{abstract}

% PACS, the Physics and Astronomy Classification Scheme.
% 03.75.Kk Dynamic properties of condensates; collective and hydrodynamic
%   excitations, superfluid flow
% 03.75.Mn Multicomponent condensates; spinor condensates
% 05.30.Jp Boson systems (for static and dynamic properties of Bose-Einstein
%  condensates, see 03.75.Hh and 03.75.Kk)
% 45.70.Qj Pattern formation of granular systems
% 47.54.+r Pattern selection; pattern formation of rarefied gas dynamics
% 42.60.Jf Beam characteristics: profile, intensity, and power;
%   spatial pattern formation of random laser
% 89.75.Kd Patterns of complex systems
\maketitle

%\section{Introduction}

Spatial domains or pattern formation is a common feature of nonlinear dynamics
in extended systems. It has been actively researched in nonlinear optics
\cite{AgrawalBook89}, classical fluids \cite{Gollub99}, granular materials
\cite{Pooley04}, and recently in atomic Bose-Einstein condensates
(BECs) \cite{Strecker02, Lewandowski02, Stenger98, Sarlo04, Kasamatsu04,
Kourakis05, Robins01}. It is generally understood that the unstable modes
of a dynamically unstable system can grow exponentially and eventually lead to
the appearance of spatial domain structures that last for a long time.

Many earlier studies have suggested interesting mechanisms for
spontaneous domain formation in atomic condensates \cite{Strecker02,
Lewandowski02, Stenger98, Sarlo04}. Most focus on the single- or
two-component condensates, where the number(s) of atoms for each component is
conserved. The dynamical instability due to
attractive atomic interactions is the most prominent among
all proposed scenarios for domain formation \cite{Strecker02, Lewandowski02,
Kasamatsu04}. The attractive interaction in a single-component condensate is
also believed to be responsible for the formation of a train of solitons,
consistent with the fact that it is dynamically unstable \cite{Strecker02}. For a
two-component condensate, again it is found that an effective attractive
interaction is responsible for the dynamical instability and domain formation
\cite{Lewandowski02, Kasamatsu04, Kourakis05}.

Several groups have also studied three-component, or spin-1 condensates
($F=1$), which are distinct as the spin mixing interaction \cite{Stenger98,
Robins01,Saito05} allows for exchanging atoms among spin components
$2|m_F=0\rangle \leftrightarrow |m_F=+1\rangle + |m_F=-1\rangle$ (hereafter as
$|0\rangle$, $|+\rangle$, and $|-\rangle$). The number of atoms for each
component therefore can change, but the total number of atoms and the system
magnetization are conserved. Significant interest now exists for spin-1
condensates because of the recent progress from several experimental groups
\cite{Barrett01}, in particular, the observation of spontaneous domain
formation in $^{87}$Rb condensates \cite{Chapman05}. Robins {\it et al.} were
among the first to study dynamical instability in a spin-1 condensate
\cite{Robins01}. They discovered a particular type of stationary state
dynamically unstable for ferromagnetic interactions, evidenced by the sudden
collapse when propagated with Gross-Pitaevskii (GP) equations, presumably
resulting from the amplification of numerical discretization errors. Through
extensive numerical simulations, Saito and Ueda also investigated very recently
the spontaneous multi-domain formation induced by the dynamical instability in
a spin-1 condensate with ferromagnetic interactions \cite{Saito05}. A clear
picture, however, is still lacking as indicated by the general lack of
comparisons with experimental reports. Our work aims at providing a complete
understanding for domain formation in a spin-1 condensate.

To begin with, we consider a homogeneous condensate at an off-equilibrium state
initially. For example, a spin-1 condensate in the ground state at a certain
nonzero magnetic (B-) field for time $t<0$ will become off-equilibrium when the
external B-field is changed for $t\ge 0$. This causes the spin-1
condensate to collectively oscillate analogous to a nonrigid pendulum as we
recently showed \cite{Zhang05a}. In addition, we assume the condensate size is
much larger than the spin healing length at least in one direction so that
domains may be formed. Within our mean field description, the evolution of a
spin-1 condensate is described by the coupled GP equations \cite{Ho98}
\begin{eqnarray}
i\hbar {\partial \over \partial t}\Phi_\pm &=&
    \left[{\cal H}
    + c_2(n_\pm+n_0-n_\mp)\right]\Phi_\pm+c_2\Phi_0^2\Phi_\mp^*,\nonumber\\
i\hbar {\partial \over \partial t}\Phi_0 &=&
    \left[{\cal H}
    + c_2(n_++n_-)\right]\Phi_0+2c_2\Phi_+\Phi_-\Phi_0^*,\label{eq:gpe}
\end{eqnarray}
where ${\cal H}=-{(\hbar^2/ 2{\sf m})}\nabla^2+V_{\rm ext}+c_0n$,
$\Phi_j$ is the j-th spin component condensate wave function, and $n_j=|\Phi_j|^2$.
$c_0=4\pi\hbar^2(a_0+2a_2)/3{\sf m}$ and $c_2=4\pi\hbar^2(a_2-a_0)/3{\sf m}$
with $a_0$ and $a_2$ the scattering lengths for the two colliding atoms
in the symmetric channels of total spin $0$ and $2$,
respectively. The interaction is ferromagnetic (antiferromagnetic) if $c_2<0$ ($>0$).

Let $\Phi_j=\sqrt {n_j}e^{i\theta_j}$ and define a relative phase $\theta =
\theta_++\theta_--2\theta_0$, Eqs. (\ref{eq:gpe})
simplifies to the following \cite{Zhang05a}
\begin{eqnarray}
\dot n_0 &=& {2c_2\over \hbar}n_0 \sqrt{(n-n_0)^2-m^2} \sin \theta,
\label{eqn0}\\
\dot \theta &=& {2c_2\over \hbar}\left[(n-2n_0)
    +{(n-n_0)(n-2n_0)-m^2 \over \sqrt{(n-n_0)^2-m^2}} \cos\theta\right], \nonumber
\end{eqnarray}
due to the conservation of
atomic density ($n=n_++n_0+n_-$) and the magnetization ($m=n_+-n_-$).
Equations (\ref{eqn0}) defines an energy conserving dynamics,
with the effective energy per unit of volume given by
\begin{eqnarray}
{\cal E}={E\over V} &=&{1\over 2}c_0n^2 + {1\over 2}c_2
\left[m^2+2n_0(n-n_0)\right. \nonumber \\
&& \left.+ 2n_0\sqrt{(n-n_0)^2-m^2}
\cos\theta \right]. \label{eq:he} % Homogeneous Energy
\end{eqnarray}
We note Eq. (\ref{eqn0}) for a homogeneous condensate differs from a trapped
one even under single spatial mode approximation despite sharing the same
dynamical Eq. (\ref{eqn0}).

%\begin{figure}
%\includegraphics[width=1.5in]{SpinDomain_Fig1.eps}
%\caption{Iso-energy contour plot of spin-1 $^{87}$Rb condensate with $m=0$
%(left) and the total spin $f$.} \label{fig1}
%\end{figure}
%\begin{figure}
%\includegraphics[width=1.5in]{./SpinSlice.eps}
%\caption{(Color) The dependence of the total spin $f$ of a homogeneous spin-1
%$^{87}$Rb condensate on the magnetization $m$, the relative phase $\theta$, and
%the $|0\rangle$ component fraction $\rho_0$. $f$ ranges from 0 to 1 as denoted
%by the legend to the right.} \label{fig:ss}
%\end{figure}

%It is easy to find that the off-equilibrium is periodic and the period is
%generally \cite{Pu99, Zhang05}
%\begin{eqnarray}
%T=.
%\end{eqnarray}
Within the mean field approximation, the average spin of a condensate
$\vec f = f_x \hat x + f_y \hat y + m \hat z$, where
$f_j=\langle F_j\rangle$ with $F_{x,y,z}$ being the
spin-1 matrices,
is also conserved in addition to the conservations of $n$ and $m$ \cite{Pu99}.
The energy functional Eq. (\ref{eq:he}) thus becomes
$
{\cal E} = {1\over 2}c_0n^2 + {1\over 2}c_2 f^2,
$
if $f^2\equiv f_x^2+f_y^2+m^2$. We note that the mean field theory
model cannot be applied to extreme cases
such as $N_0=0 $ and $N_0=N$, where quantum effects are important.

We adopt three approaches to study dynamical stability of the off-equilibrium
collective oscillations of a condensate: the effective potential method; the
Bogoliubov method; and direct numerical simulations. By going into a rotating
frame, an entire orbit reduces to a stationary point in the phase space
\cite{Zhang05a}. The effective potential then becomes ${\cal F} = (c_0/2)n^2 +
(c_2/2) \left(m^2+f_x^2+f_y^2\right)-\mu n-\eta m-\delta_xf_x-\delta_yf_y$,
where parameters $\{\mu, \eta, \delta_x, \delta_y\}=\{c_0n, c_2m, c_2f_x,
c_2f_y\}$ defines the rotating frame and are obtained through
\begin{eqnarray}
{\partial {\cal F} \over \partial n} = 0,\;\; {\partial {\cal F} \over
\partial m} = 0,\;\; {\partial {\cal F} \over \partial f_x} &=& 0,\;\;
{\partial {\cal F} \over \partial f_y} = 0. \nonumber
\end{eqnarray}

Our system is dynamically stable if its Hessian matrix of ${\cal F}$ with
respect to $\{n, m, f_x, f_y\}$ is positive definite and dynamically unstable
if the Hessian matrix has any negative eigenvalue. It is easy to check that the
eigenvalues of the Hessian matrix are $\{c_0, c_2, c_2, c_2\}$. Thus
an antiferromagnetically
interacting spin-1 condensate is dynamically stable,
while a ferromagnetically interacting one is
dynamically unstable since $c_2<0$.

We next employ the Bogoliubov transformation to find out the corresponding
unstable modes. Starting from the stationary point in the rotating frame as found
above, the equation of motion for collective excitations
can be cast in a matrix form \cite{Maldonado93} as $ {\cal M}\cdot
\vec x = \hbar \omega \vec x$, with a vector $\vec x =
(\delta\Psi_+, \delta\Psi_0, \delta\Psi_-, \delta\Psi_+^*, \delta\Psi_0^*,
\delta\Psi_-^*)^T$. $\delta\Psi_j$ and $\delta\Psi_j^*$ denote the
deviations from the stationary point, and the
associated matrix is
\begin{eqnarray}
{\cal M} &=& \left(\begin{array}{cc}{\cal A}&{\cal B}\\-{\cal B}^*&-{\cal
A}^*\end{array}\right), \nonumber
\end{eqnarray}
with
\begin{widetext}
\begin{eqnarray}
{\cal A} &=& \left(\begin{array}{ccc}\varepsilon_k + (c_0+c_2)n_+ + c_2n_0
&c_0\Phi_0^* \Phi_+ + c_2\Phi_0\Phi_-^*
&(c_0-c_2) \Phi_-^* \Phi_+ \\
c_0\Phi_0 \Phi_+^* + c_2\Phi_0^*\Phi_- &\varepsilon_k + c_0n_0+ c_2(n_++n_-)
&c_0\Phi_0 \Phi_-^* + c_2\Phi_0^* \Phi_+ \\
(c_0-c_2) \Phi_- \Phi_+^* &c_0\Phi_0^* \Phi_- + c_2\Phi_0 \Phi_+^*
&\varepsilon_k + (c_0+c_2)n_- + c_2n_0\end{array}\right), \nonumber
\end{eqnarray}
and
\begin{eqnarray}
{\cal B} &=& \left(\begin{array}{ccc}(c_0 + c_2) \Phi_+^2 &(c_0 + c_2) \Phi_0
\Phi_+
&c_2\Phi_0^2 + (c_0 - c_2)\Phi_-\Phi_+\\
(c_0 + c_2) \Phi_0 \Phi_+ &c_0\Phi_0^2 + 2c_2\Phi_-\Phi_+
&(c_0 + c_2) \Phi_0\Phi_-\\
c_2\Phi_0^2 + (c_0 - c_2)\Phi_-\Phi_+ &(c_0 + c_2) \Phi_0\Phi_- &(c_0
+c_2)\Phi_-^2\end{array}\right). \nonumber
\end{eqnarray}
\end{widetext}
$\varepsilon_k = \hbar^2 k^2/2{\sf m}$ is kinetic energy of the collective
excitation mode with wave vector $k$.

\begin{figure}
\begin{center}
\includegraphics[width=3.25in]{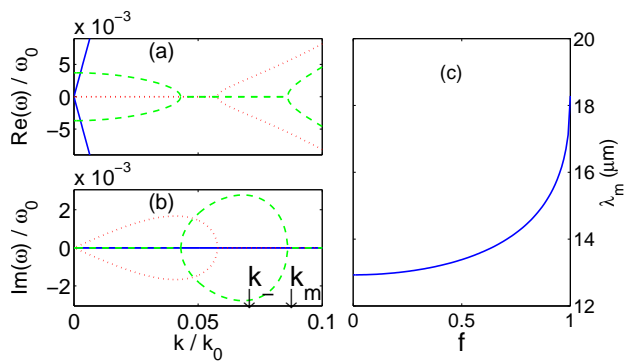}
\caption{(Color online) Real part (a) and imaginary part (b) of a typical
Bogoliubov spectrum for a spin-1 condensate with ferromagnetic interaction.
Panel (c) shows the smallest wavelength of a homogeneous spin-1 $^{87}$Rb
condensate at $n=1.9\times 10^{14}$ cm$^{-3}$.} \label{fig:ee}
\end{center}
\end{figure}

The eigen-frequencies of the Bogoliubov excitations are obtained
from the characteristic equation $\det({\cal M}-\hbar \omega I)=0$,
explicitly given by
\begin{eqnarray}
\left[2c_s\varepsilon_k + \varepsilon_k^2+c_s^2{\sf
f}^2-(\hbar\omega)^2\right]&&
\nonumber \\
\times \left[\left(\varepsilon_k^2-(\hbar\omega)^2\right)
\left(2c_s\varepsilon_k+\varepsilon_k^2-(\hbar\omega)^2\right)\right. &&\nonumber \\
\left.+2c_n\varepsilon_k\left(\varepsilon_k^2+2c_s\varepsilon_k(1-{\sf
f}^2)-(\hbar\omega)^2\right)\right] &=& 0,
\end{eqnarray}
with $c_n=c_0n$, $c_s=c_2n$, and ${\sf f}=f/n$.
The frequencies are then given by
\begin{eqnarray}
(\hbar\omega)^2_{1,2} &=&
\varepsilon_k\left[(c_n+c_s+\varepsilon_k)+\sqrt{(c_n-c_s)^2+4c_nc_s{\sf
f}^2}\,\right],
\nonumber \\
(\hbar\omega)^2_{3,4} &=&
\varepsilon_k\left[(c_n+c_s+\varepsilon_k)-\sqrt{(c_n-c_s)^2+4c_nc_s{\sf
f}^2}\,\right],
\nonumber \\
(\hbar\omega)^2_{5,6} &=& (\varepsilon_k+c_s)^2-c_s^2 (1-{\sf f}^2).
\label{eq:ee}
\end{eqnarray}
The corresponding modes are termed as density modes (solid lines),
spin modes (dotted lines), and quadrupolar spin modes (dashed lines as in Fig. \ref{fig:ee})
by Ho \cite{Ho98}. Figure \ref{fig:ee}(a) and
\ref{fig:ee}(b) show the real and imaginary parts of the typical
dispersion relation for a $^{87}$Rb spin-1 condensate \cite{para2},
respectively. All frequencies are real for
antiferromagnetically interacting (e.g. $^{23}$Na) condensate.

Our analysis here parallels that of Refs.
\cite{Kasamatsu04,Kourakis05} for a two-component condensate.
We find two interesting features in Fig. \ref{fig:ee}(b).
One of them at $k_-$ is the most unstable mode at
the maximum imaginary frequency,
it determines short
time behavior such as the time scale for domains to emerge. The
other is the largest wave vector $k_m$ with an (infinitesimal)
imaginary frequency which determines the long time behavior such
as the final domain size. From Eq. (\ref{eq:ee}), we find
the time scale for the emergence of a domain is $\sim
1/|\omega(k_-)| = h/(|c_s|\sqrt{1-{\sf f}^2})$ and the domain
width is about $\lambda_m = 2\pi/k_m = h /
\sqrt{2m|c_s|(1+\sqrt{1-{\sf f}^2})}$.
It is typically of the order of spin healing length.
Figure \ref{fig:ee}(c)
displays the ${\sf f}$-dependence of $\lambda_m$ at $n=1.9\times
10^{14}$ cm$^{-3}$ for a homogeneous $^{87}$Rb spin-1 condensate.
For the domains to form, a
condensate has to be larger than $\lambda_m$, at least in one direction.

We note the spin domain formation as discussed here is different
from striation patterns as observed in (antiferromagnetic)
$^{23}$Na condensates. The stripe patterns arise from interplay of
an external B-field, a field gradient, and immiscibility
among different spin components \cite{Stenger98}. No domains were
observed in Stenger {\it et al.}'s experiment \cite{Stenger98} at
negligible B-fields where the $|+\rangle$ and $|-\rangle$
components coexist. At finite values of B-fields, phase
separation between the $|0\rangle$ and the $|+$/$-\rangle$
components occurs \cite{Zhang03}. In Miesner {\it et al.} and
Stamper-Kurn {\it et al.}'s experiments \cite{Stenger98},
only two spin components were involved due to the relatively large
bias B-field ($\sim 15$ G).

\begin{figure}
\includegraphics[width=3.25in]{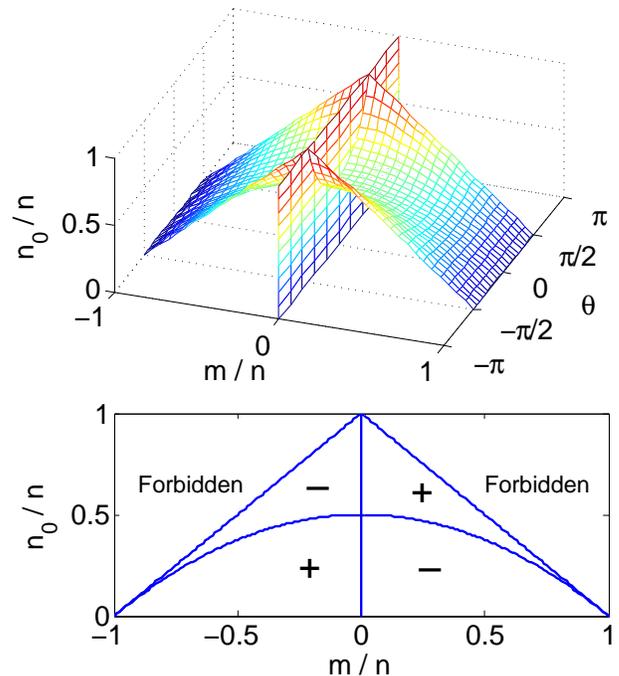}
\caption{(Color online) Surfaces of $d{\cal E}/dm=0$ (top) and the
cross-section at $\theta=0$ (bottom). The plus signs denote $d{\cal E}/dm>0$
and the minus signs denote $d{\cal E}/dm<0$.}
\label{fig:dfdi} % Domain Formation from Dynamical Instability
\end{figure}

Our analysis shows the formation of spin domains is
a direct consequence of dynamic instability for a condensate with
ferromagnetic interactions. To provide a clearer physical picture
for domain formation,
we now work in the lab frame. We focus on $d{\cal E}/dm$,
which in fact calibrates the formation of the spin domain. We find
\begin{eqnarray}
{d{\cal E} \over dm} &=& c_2m\left[ 1 -
    {n_0\cos\theta\over \sqrt{(n-n_0)^2-m^2}} \right].
\end{eqnarray}
Figure \ref{fig:dfdi} shows the surfaces where the above first order derivative
is zero. The region below the saddle surface in Fig. \ref{fig:dfdi} of an orbit
is unstable if $c_2<0$. Here the meaning of ``unstable" is generalized,
referring to the dynamical property where the local magnetization tends to
deviate further from $m=0$. For example, in the lower right allowed region of
$d{\cal E}/dm < 0$, $\Delta m>0$ is required to lower the local energy ${\cal
E} \approx {\cal E}(m) + (d{\cal E}/dm) \Delta m$. Thus $m$ tends to increase.
Similarly the lower left allowed region would make $m$ decrease. The combined
effect is dynamically unstable orbits, separation of $|+\rangle$ and
$|-\rangle$ component, and the eventual formation of spin domains.

For antiferromagnetic interactions ($c_2>0$), $\theta$
usually oscillates around $\pi$. Thus $d{\cal E}/dm>0$ for $m>0$ and $d{\cal
E}/dm<0$ for $m<0$. So the magnetization always oscillates around zero and
no domain forms. This coincides with the findings of the dynamical stability
analysis for an antiferromagnetically interacting condensate.

\begin{figure}
\includegraphics[width=3.25in]{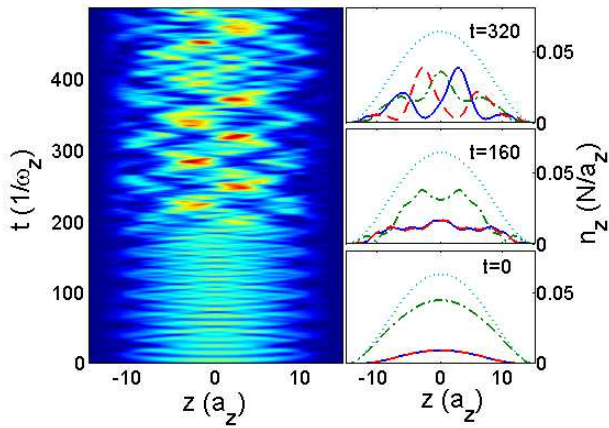}
\caption{(Color online) Typical evolutions for spin domain formation in a
$^{87}$Rb condensate. The initial state is the
ground state at $B=0.3$ G. The B-field is then set to zero and a small
white noises are added throughout the evolution. The left contour plot is
for the $|+\rangle$ component. The right column shows the density
distribution of all three components at times $t=0,160,320$ ($1/\omega_z$).
Solid, dash-dotted, and dashed lines denote respectively the $|+\rangle$,
$|0\rangle$, and $|-\rangle$ component. Dotted lines are for the total density.
The axial density is $n_z\equiv\int \sum_j|\Phi_j|^2 2\pi r dr$.
$a_z=\sqrt{\hbar/{\sf m}\omega_z}\simeq 2.2\mu$m and the average condensate
density is $\sim 1.9\times 10^{14}$ cm$^{-3}$.}
\label{fig:sd} % Spin Domains
\end{figure}

Finally we perform numerical simulations of Eq. (\ref{eq:gpe}) to confirm the
mechanism of dynamical instability-induced spin domain formation. The initial
conditions are as in the experiment \cite{Chapman05}, with $^{87}$Rb
condensates ($N_0(0)/N=0.744$, $\theta(0)=0$, for the ground state of
$N=2.0\times 10^5$ at $B=0.3$ G and $M=0$), in a trap $V_{\rm ext}(\vec r) =
({\sf m}/2)(\omega_x^2x^2 +\omega_y^2y^2+\omega_z^2z^2)$ with
$\omega_x=\omega_y= (2\pi) 240$ Hz and $\omega_z=(2\pi) 24$ Hz. In one of the
simulations, we intentionally include additive small white noise ($\sim
1.0\times 10^{-5}$), although still much larger than numerical errors
\cite{Kramer04} during the propagation. We find that it takes a shorter time
for the $|+\rangle$ and $|-\rangle$ components to separate when white-noise is
included. Figure \ref{fig:sd} shows the evolution of axial density
distributions. Phase separation between the $|+\rangle$ and $|-\rangle$
components is seen accompanied by the formation of domains. This proves again
that dynamical instability causes the formation of domains. The domain width
(an upper limit) as estimated from $\lambda_m = {h/\sqrt{2{\sf m}|c_2|\langle
n\rangle(1+\sqrt{1-{\sf f}^2})}} \approx 15 \mu$m, is consistent with both
simulations and experimental observations \cite{Chapman05}.

In a spinor condensate, spin wave excitations normally refer to
dynamically stable (or relatively more stable) collective modes.
Once excited, they lead to coherent cyclic dynamics in both
spatial and temporal dimensions. Spin domains, on the other hand,
refers to unstable modes, with a fixed pattern in the long time limit.

In conclusion we have presented a systematic study of dynamical stability and
the accompanied mechanism for domain formation in a spin-1 condensate. Our
results affirm that a ferromagnetically interacting condensate is dynamically
unstable and evolves spontaneously into multi-domain structures contrary to
dynamically stable antiferromagnetic condensates. Our work provides
a clear physical picture for recently observed spontaneous domain formations in
spin-1 condensates \cite{Chapman05}.

This work is supported by NASA and NSF.

\end{document}